\documentclass[preprint]{mn2e}

\usepackage{epsfig}

\newcommand{\etal}{et~al.\ }
\newcommand{\eg}{e.g.\ }
\newcommand{\ie}{i.e.\ }
\newcommand{\Msun}{M_{\odot}}

\newcommand{\SiII}{Si~{\sc ii}}

\newcommand{\SII}{S~{\sc ii}}

\newcommand{\TiII}{Ti~{\sc ii}}
\newcommand{\CrII}{Cr~{\sc ii}}

\newcommand{\FeII}{Fe~{\sc ii}}
\newcommand{\FeIII}{Fe~{\sc iii}}
\newcommand{\CoII}{Co~{\sc ii}}

\newcommand{\NiII}{Ni~{\sc ii}}
\newcommand{\Feff}{$^{54}$Fe}
\newcommand{\Fefs}{$^{56}$Fe}
\newcommand{\Cofs}{$^{56}$Co}
\newcommand{\Nifs}{$^{56}$Ni}
\newcommand{\Nife}{$^{58}$Ni}
\newcommand{\Mej}{$M_{\rm ej}$}

\newcommand{\DeltaB}{$\Delta m_{15}(B)$}
\newcommand{\DeltaV}{$\Delta m_{15}(V)$}
\newcommand{\DeltaBol}{$\Delta m_{15}({\rm Bol})$}
\newcommand{\kopt}{\kappa_{\rm opt}}
\def\lsim{\mathrel{\rlap{\lower 4pt \hbox{\hskip 1pt $\sim$}}\raise 1pt\hbox {$<$}}}
\def\gsim{\mathrel{\rlap{\lower 4pt \hbox{\hskip 1pt $\sim$}}\raise 1pt\hbox {$>$}}}

\title{The (\Feff+\Nife)/\Nifs\ ratio as a second parameter for Type Ia 
  supernova properties}

\author[Mazzali \mbox{\rm\&} Podsiadlowski]
{Paolo A.~Mazzali$^{1,2}$\thanks{E-mail: mazzali@mpa-garching.mpg.de},
Philipp\ Podsiadlowski$^3$\\
$^{1}${\it Max-Planck Institut f\"ur Astrophysik,
  Karl-Schwarzschildstr. 1, 85748 Garching, Germany}\\
$^{2}${\it Istituto Naz. di Astrofisica-Oss.\ Astron., Via Tiepolo, 11,
  34131 Trieste, Italy}\\
$^{3}${\it Dept.\ of Astronomy, Oxford University, Oxford, OX1 3RH, UK}
}
\date{\today}

\volume{000}

\pagerange{1\,--\,7} \pubyear{2006}

\begin{document}
\maketitle

\label{firstpage}

\begin{abstract}

A variation of the relative content of (\Feff+\Nife) versus \Nifs\ may
be responsible for the observed scatter of Type Ia Supernovae (SNe Ia)
about a mean relation between their intrinsic brightness and the shape
of their light curve. Synthetic light curves are computed of
parametrised Chandrasekhar-mass explosion models of constant kinetic
energy, where the ejecta are divided into an inner NSE zone, composed
of (\Feff+\Nife) inside and \Nifs\ outside, an outer zone with
Intermediate Mass Elements and a CO zone.  Both the size of the NSE
zone and the fraction of (\Feff+\Nife) v. \Nifs\ are varied
systematically.  Models with the same original NSE content but
different (\Feff+\Nife)/\Nifs\ ratios reach different peak brightness
but have similar light curve shapes.  Synthetic spectra indicate that
the $V$-band decline rate is not affected by the (\Feff+\Nife)/\Nifs\
ratio.  While the \Nifs\ mass and the total NSE mass are the dominant
parameters determining the peak luminosity and the shape of the light
curve, respectively, a variation in the (\Feff+\Nife)/\Nifs\ ratio,
which may depend on the metallicity of the progenitor (Timmes, Brown
\& Truran 2003) is likely to account for a significant part of the
observed scatter of local SNe Ia about the mean brightness--decline
rate relation.

\end{abstract}



\section{Introduction}

The use of Type Ia Supernovae (SNe Ia) to constrain cosmological
parameters relies on the calibration of their luminosity.  Although
SNe Ia are not standard candles, several of their observed properties
can be linked to their peak brightness.  These properties include the
decline in $B$-band magnitude in the first 15 days after $B$ maximum,
\DeltaB\ (Phillips 1993; Phillips et al.\ 1999), the shape of the
light curve (\eg MLCS (Riess et al.\ 1998), ``stretch'' (Perlmutter et
al.\ 1997)), the ratio of two \SiII\ lines (Nugent et al.\ 1995), the
width of the nebular emission lines (Mazzali et al.\ 1998), the Hubble
type of the parent galaxy (Hamuy et al.\ 2000; Branch, Romanishin \&
Baron 1996; van den Bergh 1997; van den Bergh, Li \& Filippenko 2005),
the distance from the centre of the host galaxy (Wang, H\"oflich \&
Wheeler 1997), and the $B-V$ colour $\sim 12$ days after $B$ maximum
(Wang et al.\ 2005).

All these relations indicate that the properties of SNe Ia are
determined by one dominant, driving parameter, which should be
closely related to the amount of \Nifs\ synthetised in the explosion
(H\"oflich et al.\ 1996). However, scatter about the mean relation is
always present. This scatter may be due to uncertainties in distance
and reddening affecting the estimate of the SN brightness, but it
remains even when distance- and reddening-free quantities are used
(e.g., nebular line width vs.\ \DeltaB; Mazzali et al.\ 1998). Apart
from possible uncertainties in the measurements, this scatter suggests
that a single parameter is not sufficient to describe all possible
variations of the SN~Ia phenomenon. This has also been confirmed by
detailed studies of individual SNe (Benetti et al.\ 2004).

The observed scatter could be due to a number of factors, that may in turn be
related to properties of the progenitor system (\eg distribution of \Nifs\ in
the ejecta, mass and composition of the exploding white dwarf). The
identification of a possible second parameter may help us understand how SNe~Ia
explode and improve the calibration of their luminosity.  Here we focus on the
possible role of a variation of the ratio $R =$(\Feff+\Nife)/\Nifs, following a
suggestion by Timmes et al.\ (2003).

\section{Light curve properties and the role of the (\Feff+\Nife)/\Nifs\ 
ratio}

The peak brightness of a SN~Ia is proportional to the mass of \Nifs\
synthesized (Arnett 1982, 1996). In fact, the decay of \Nifs\ into
\Cofs\ and \Fefs\ produces $\gamma$-rays and positrons that deposit
their energy in the SN ejecta and eventually cause it to shine.  It
also follows from simple dimensional analysis that the light curve
width $\tau_{\rm LC}$ depends on the ejected mass $M_{\rm ej}$, the
kinetic energy of the explosion $E_{\rm K}$, and the grey opacity to
optical photons $\kopt$ as
\begin{equation}
\tau_{\rm LC} \propto \kopt^{1/2} M_{\rm ej}^{3/4} E_{\rm K}^{-1/4}.
\end{equation}
If the white dwarf mass and the explosion energy are assumed to be
constant, the shape of the light curve must depend primarily on the
opacity. It is now understood (Khokhlov, M\"uller \& H\"oflich 1993;
H\"oflich et al.\ 1996) that the two quantities, mass of \Nifs\ and
opacity, are intimately related, so that SN~Ia light curves 
behave to first order as a one-parameter family.

The thermonuclear explosion that gives rise to a SN~Ia (e.g. Nomoto et
al.\ 1984) is thought to produce material in nuclear statistical
equilibrium (NSE) in the inner parts of the white dwarf. Most of this
should be iron-peak nuclei. In the innermost $\sim 0.2\Msun$, where the
density is high enough for weak interactions to be important, \Nife\ and
\Feff\ are mainly produced. In mass shells between $\sim 0.2\Msun$ and
$\sim 0.8\Msun$, weak interactions operate on a timescale long compared
to the nuclear burning timescale, so that the electron fraction $Y_e$
remains constant in this region during the burning phase; as a
consequence, the final relative abundances of \Nife, \Nifs, and \Feff\
in this layer are directly related to the value of $Y_e$ at the time of
the explosion. The amount of \Nifs\ produced is essential in determining
the brightness of the SN.

Further out still, the densities are too low to burn to NSE, and
intermediate-mass elements (IME) such as Si and S are produced. While
no $\gamma$-rays and positrons are generated here, burning to Si
produces almost as much kinetic energy as burning to NSE (Gamezo,
Khokhlov \& Oran 2005).  Therefore this region is also important in
determining the properties of the SN. Finally, an outer region
consisting of unburned C and O may also exist.

Producing a larger amount of \Nifs\ leads to more heating. Khokhlov et
al.\ (1993) and H\"oflich et al.\ (1996) showed that the opacity
increases rapidly with temperature for $T < 10^4$\,K, the effective
temperature of a typical SN~Ia near maximum, but is relatively
insensitive to $T$ for higher values. Moreover, since most \Nifs\ is
not mixed, a larger mass of \Nifs\ does not immediately translate into
a higher temperature outside the \Nifs\ zone.


The explanation of this behaviour lies in the fact that, since
continuum processes are depressed in the ejecta of a SN~Ia because of
the lack of hydrogen and helium, the opacity is dominated by line
opacity of low ionisation species (Pauldrach et al.\ 1996). Combined
with the large differential motion of the SN envelope, line absorption
causes a pseudo-continuum opacity which is essential in SN~Ia light
curves. Owing to their complex atomic level structure, ions of
Fe-group elements (\eg \FeII, \FeIII, \NiII, \CoII, \TiII, \CrII) have
many more active lines than IME ions (\eg \SiII, \SII) and make the
dominant contribution to the line opacity (Mazzali et al.\ 2001).

Therefore, while producing more \Nifs\ means producing more photons
(\ie a brighter maximum), the opacity and the shape of the light curve
depend on the total amount of NSE material synthesised. The line
opacity is in fact the same in different isotopes of the same element,
(\eg \Fefs\ and \Feff).  Since brighter SNe~Ia have broader light
curves, the implication is that they do not only make more \Nifs, but
also more NSE material. This suggests that brighter SNe do not simply
make more \Nifs\ at the expense of \Nife\ or \Feff.

Mazzali et al.\ (2001) built parametrised explosion models where the
total NSE plus IME content is constant, so that $E_{\rm K}$ can also be
regarded as constant, but different amounts of \Nifs\ are synthesised at
the expense of IME. This was achieved by moving the outer boundary of
the \Nifs\ zone. Synthetic bolometric light curves of these models
reproduced the observed $M_{\rm Bol}$(Max)--\DeltaB\ relation (Contardo,
Leibundgut \& Vacca 2000). Using spectrum synthesis, Mazzali et al.\
(2001) showed that the relation between $M_B$(\rm Max) and \DeltaB\ can
also be recovered.

The role of \Nife\ and \Feff\ is only to provide an opaque inner
zone. Since this zone is very well hidden from the outer layers, the
effect of its presence on the early light curve is limited. In this
paper we explore the effect of different $R$ values for the same total
mass of NSE material.  Timmes et al.\ (2003) estimate that the final
\Nifs\ mass fraction depends on the initial metallicity $Z$ as
$X$(\Nifs)$\approx 1 - 0.054\,Z/Z_{\odot}$, a relation confirmed by
detailed three-dimensional explosion and nucleosynthesis calculations
by R\"opke et al.\ (2005) and Travaglio, Hillebrandt \& Reinecke
(2005).  Both H\"oflich, Wheeler \& Thielemann (1998) and Dominguez,
H\"oflich \& Straniero (2001), using delayed detonation models, also
suggest that $R$ increases with the metallicity of the
progenitor. They find that their monochromatic and bolometric light
curves are very little affected by changes in $R$.  We explore the
dependence in a more systematic way, independent of a specific model
of the explosion. Therefore, we consider a situation where \Nife\ and
\Feff\ are present not only at the lowest velocities, where they
dominate, but also in the \Nifs\ zone itself.

\section{Light Curve Modelling}

Mazzali et al.\ (2001) built parametrised explosion models consisting
simply of 3 zones: the innermost zone (\Feff, \Nife) only contributes
to the opacity. Outside of this, a \Nifs\ zone produces $\gamma$-rays
and makes a large contribution to the opacity.  The outer zone (IME
plus CO) only makes a small contribution to the opacity.  They studied
three representative explosion models. The models have the same \Mej\
($1.4 \Msun$) and $E_{\rm K}$ ($\sim 1.35\times 10^{51}$\,erg), but
contain different amounts of \Nifs: 0.4, 0.6, and $0.8\Msun$,
respectively, and were accordingly called Ni04, Ni06, and Ni08.  They
reproduced the observational properties of SNe 1992A, 1994D, and
1990N, which span the range of spectroscopically normal SN~Ia decline
rates.  Starting from those models, we introduced \Nife\ and \Feff\ in
the \Nifs\ layer in two different amounts, 10\% and 20\% relative to
\Nifs, and with the same distribution, following the predictions of
Timmes et al.\ (2003), and computed synthetic bolometric light curves.

These were computed with the gray Montecarlo code developed by Mazzali
et al.\ (2001).  The code follows the propagation and deposition of the
$\gamma$-ray photons and the positrons emitted in the decay of \Nifs\ to
\Cofs\ and hence to \Fefs. It then follows, also in Montecarlo, the
diffusion of the optical photons until they emerge to give rise to the
observable SN light. The code adopts a simple parametrisation of the
line opacity, based on the number of active lines in different elements.
In particular, it separates the effect of NSE elements from that of IME
and of the original CO of the white dwarf.  It also takes into account
the decrease of the opacity after maximum caused by the cooling of the
ejecta, as described by Khokhlov et al.\ (1993) and H\"oflich et al.\
(1996). The opacity is parametrized as follows:
\begin{equation} 
	\kappa_{\rm opt} = \left[ 0.25 X_{\rm Fe} + 
	0.025 (1 - X_{\rm Fe} )\right] \left( \frac{t_d}{17}
	\right)^{-\frac{3}{2}} [\rm cm^2 g^{-1}], 
\end{equation} 
where $X_{\rm Fe}$ is the mass fraction in Fe-group elements and $t_d$ the time
since the explosion in days. The time-dependent term mimicks the effect of the
decreasing temperature and is limited to a maximum value of 2.

Since the \Nifs\ mass of the modified models is smaller by 10\% and 20\%,
respectively, with respect to that of the original ones, it can be expected that
the peak brightness of their respective light curves will be reduced
proportionally.  However, the effect on the opacity should be very small:
according to Eq.~(2), \Nife\ and \Feff\ contribute to the opacity as much as
\Nifs. In practice, we tested models with different \Nifs\ masses, but with the
same amount and distribution of opacity. The bolometric light curves thus
obtained are shown in Figure 1, and their most important properties are listed in
Table 1.

For each set of light curves in Figure 1, the original light curve is
the brightest one. As the \Nife\ and \Feff\ content increases and the
\Nifs\ content decreases, the light curves become fainter as
expected. However, since models with the same mass of NSE material
have similar opacities, their light curves have similar shapes. In
contrast, for example, the Ni08-20\% model light curve is similar to
that of the Ni06 model in peak brightness, but the two light curves
have different decline rates. Although the bolometric decline rates
are in good agreement with observed values for SNe~Ia covering the
range of normal spectral properties (Contardo et al.\ 2000), the
occurrence of SNe with different $R$ values can generate a dispersion
about the mean $M_{\rm Bol}(\rm Max)$--\DeltaBol\ relation. Note that
this would not be the case if the changes in composition had been
introduced by extending the inner (\Nife\ + \Feff) zone outwards: this
would lead to the average location of \Nifs\ moving outwards as its
mass becomes smaller and hence to a smaller photon diffusion
time. This family of light curves would therefore follow the
brightness-decline rate relation.

\section{Effect on Monochromatic Decline Rates}

Our bolometric light curves were produced using a simple code, whose main
advantage is that it makes it easy to trace the effect of properties such as
the \Nifs\ distribution and that it is not subject to uncertainties arising
from the use of opacity tables. Obviously, it would be very interesting to
verify how the monochromatic light curves are affected, as these are directly
observable. As was done in Mazzali et al.\ (2001), we checked the behaviour of
the monochromatic bands computing spectra at the appropriate epochs (maximum
and 15 days later) with our Montecarlo spectrum synthesis code (Mazzali \& Lucy
1993; Lucy 1999; Mazzali 2000). The bolometric luminosities and the
photospheric velocities obtained from our synthetic light curves were used as
input for the spectral calculations.

The results for the spectra at maximum are shown in Table 2. Interestingly,
although the luminosity of the models with a higher \Nife\ and \Feff\
content is smaller than that of the original models, the colours of the
spectra are essentially unchanged, \ie both $B$ and $V$ change by roughly
equal amounts.  This confirms the findings of H\"oflich \etal (1998) and
Dominguez \etal (2001) and provides further support to the notion that the
typical colour of normal SNe~Ia at maximum light is $(B-V) \sim 0.0$--0.1.
Furthermore, the brightest SNe (model Ni08) are not the bluest. This is due
to the changing ionisation conditions, with strong \FeIII\ lines developing
in the $B$ band.  This change from \FeII\ to \FeIII\ does not affect the
light curve much: opacity is shifted to the blue, but so is the average flux
since the temperature is higher.  Originally, the three spectra resembled
those of SNe 1992A, 1994D, and 1990N, respectively (Mazzali et al.\ 2001,
Figure 5). The spectra of the modified models are very similar, except for
an offset in flux and a slight reduction in temperature as the mass of
\Nifs\ is reduced.

The shape of the synthetic spectra 15 days after maximum depends sensitively
on the choice of photospheric velocity, which is very uncertain at such
advanced epochs. This affects in particular the $B$ magnitudes and the $B-V$
colour. However, small changes to lower velocities than those predicted by
the light curve code are sufficient to obtain spectra with \DeltaB\ values
similar to the original models.  On the other hand, the $V$ magnitude is
almost unaffected by the choice of photospheric velocity within this range.
Therefore, the value of \DeltaV\ can be determined quite accurately. We
computed \DeltaV\ rather than $\Delta m_{20}(V)$, the value measured by
Hamuy et al.\ (1996b), because the position of the photosphere would be even
more uncertain at a later epoch.  The values of \DeltaV\ for our model set
(Table 2) compare well with those of observed SNe~Ia (Hamuy et al.\ 1996b,
Table 1). Since \DeltaB\ and $\Delta m_{20}(V)$ are observationally very
well correlated (Hamuy et al.\ 1996b), the agreement between our computed
values of \DeltaV\ and the observed ones supports our hypothesis that
\DeltaB\ does also not change significantly with $R$.

Our calculations confirm that SNe with constant total NSE content but
(\Nife\ + \Feff)/\Nifs\ ratios varying within the range suggested by Timmes
et al. (2003) span a range of $\sim 0.25$ mag in peak brightness.  However,
they  have similar $V$- and $B$-band decline rates.  If SNe with a variable
$R$ do exist, they would show a scatter in the peak magnitude--decline rate
diagram comparable to the observed one. This is illustrated in Figure~2,
where we plot the location of our models in this diagram and the spread of
observed SN properties: our models are able to reproduce both the
brightness--decline relation and the observed scatter. This suggests that
$R$ may be one of the most important parameters responsible for the
dispersion in the brightness--decline relation.

More accurate pre-SN calculations and detailed explosion models may be able
to determine the value and range of the ratio (\Nife\ + \Feff)/\Nifs\
depending on properties of the progenitor white dwarf such as metallicty or
age, (cf.\ Lesaffre et al.\ 2006). At the same time, it will be important to
verify the observable effect the ratio has on the observable properties of a
SN~Ia with more accurate radiative transfer calculations. Quantifying the
effect and correcting for this may ultimately lead to a reduction of the
scatter in the peak brightness--decline rate relation. A way to separate the
first and the second parameter for nearby SNe could be the following. On the
one hand, the mass of \Nifs\ can be derived from the late light curves, when
the SN brightness is more directly sensitive to $\gamma$-ray and positron
deposition. On the other hand, nebular spectra could be used to obtain
information on both the mass of \Nifs\ and that of non-radioactive Fe-group
isotopes.



\clearpage


\begin{table*}
\caption{Parameters of the synthetic light curves.}
\begin{tabular}{lccccc}
\hline\hline
\noalign{\vspace{2pt}}
model&{$M$(\Nifs)}&$t_{\rm Max}$&$M{\rm(Bol)}_{\rm Max}$ &
$M{\rm (Bol)}_{15}$&{$\Delta m_{15}$(Bol)}\\
&$[\Msun]$&[d] &[mag] & [mag] & [mag] \\
\noalign{\vspace{2pt}}
\hline
\noalign{\vspace{2pt}}
Ni08 	   & 0.80 & 18.5 & $-19.33$ & $-18.40$ & 0.93 \\
Ni08-10\%  & 0.72 & 18.5 & $-19.21$ & $-18.28$ & 0.93 \\
Ni08-20\%  & 0.64 & 18.5 & $-19.08$ & $-18.10$ & 0.92 \\
Ni06 	   & 0.60 & 18.0 & $-19.15$ & $-18.16$ & 1.01 \\
Ni06-10\%  & 0.54 & 18.0 & $-19.03$ & $-18.05$ & 0.98 \\
Ni06-20\%  & 0.48 & 18.0 & $-18.90$ & $-17.92$ & 0.98 \\
Ni04 	   & 0.40 & 18.0 & $-18.87$ & $-17.82$ & 1.05 \\
Ni04-10\%  & 0.36 & 18.0 & $-18.76$ & $-17.71$ & 1.05 \\
Ni04-20\%  & 0.32 & 18.0 & $-18.63$ & $-17.58$ & 1.05 \\
\noalign{\vspace{2pt}}
\hline
\end{tabular}
\end{table*}


\begin{table*}
\caption{Properties of the spectra and inferred value of the decline 
rates.}
\begin{tabular}{lccccc}
\noalign{\vspace{2pt}}
\hline\hline
model&{$M(B)_{\rm Max}$} &
{$(B-V)_{\rm Max}$} &
{$(B-V)_{15}$} &
{$\Delta m_{15}(B)$}    &
{$\Delta m_{15}(V)$}     \\
&
{[mag]} &
{[mag]} &
{[mag]} &
{[mag]} &
{[mag]} \\
\noalign{\vspace{2pt}}
\hline
\noalign{\vspace{2pt}}
Ni08 	  & $-19.22$ & 0.10 & 0.56 & 1.08 & 0.62 \\
Ni08-10\% & $-19.08$ & 0.10 & 0.61 & 1.08 & 0.57 \\
Ni08-20\% & $-18.94$ & 0.16 & 0.65 & 1.09 & 0.60 \\
Ni06 	  & $-19.12$ & 0.06 & 0.70 & 1.31 & 0.67 \\
Ni06-10\% & $-19.03$ & 0.08 & 0.71 & 1.30 & 0.67 \\
Ni06-20\% & $-18.85$ & 0.08 & 0.79 & 1.32 & 0.61 \\
Ni04 	  & $-18.88$ & 0.06 & 0.81 & 1.46 & 0.71 \\
Ni04-10\% & $-18.76$ & 0.06 & 0.84 & 1.46 & 0.68 \\
Ni04-20\% & $-18.63$ & 0.07 & 0.81 & 1.45 & 0.71 \\
\noalign{\vspace{2pt}}
\hline
\end{tabular}
\end{table*}



\clearpage
\begin{figure*}
\psfig{figure=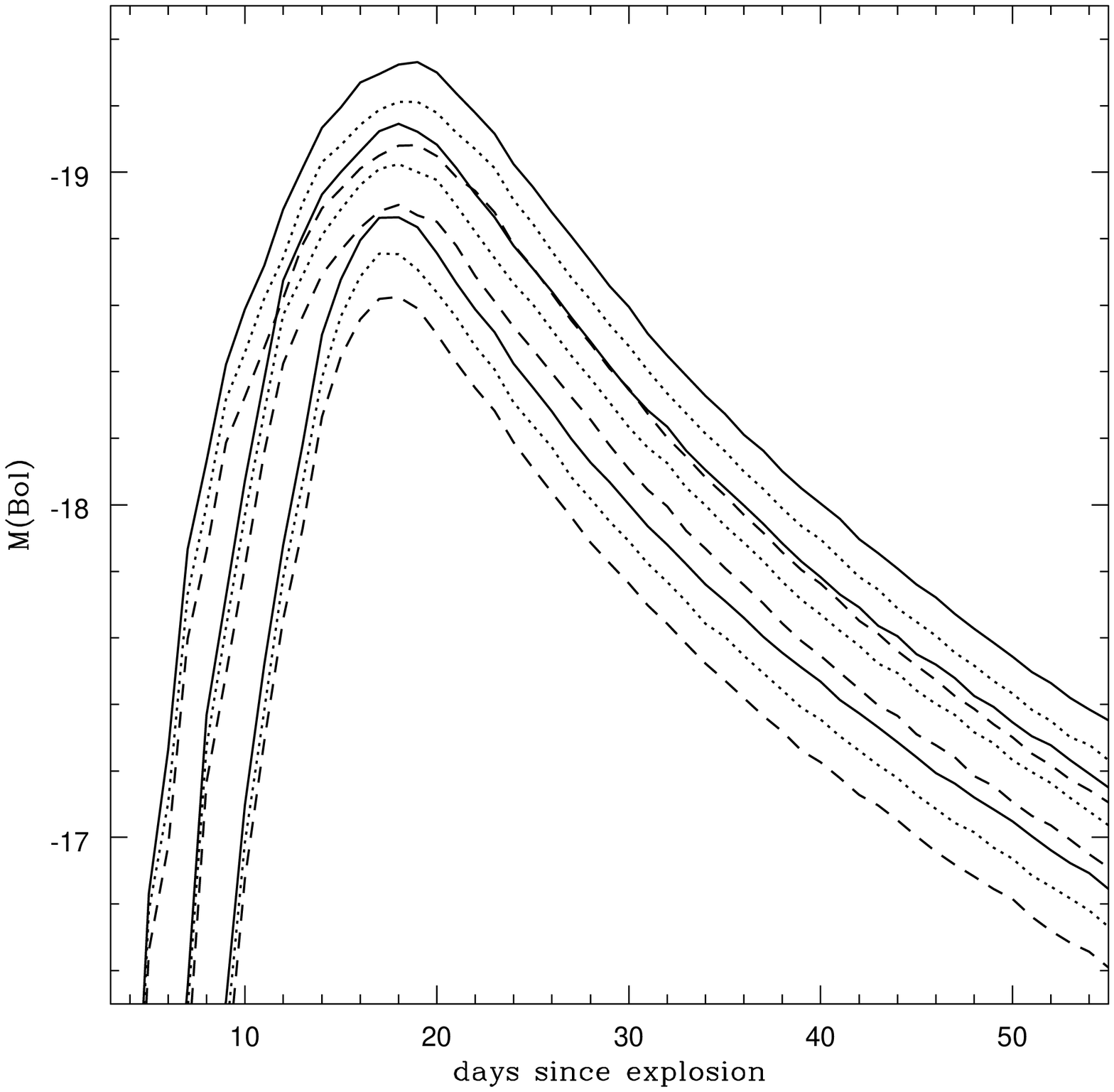,width=12cm}
\caption{Families of bolometric light curves computed for models 
Ni08, Ni06 and Ni04, with and without the inclusion of \Nife\ and \Feff.  
The continuous curves are for the original models (with $0.8\Msun$, $0.6\Msun$, 
and $0.4\Msun$ of \Nifs, respectively), the dotted ones are for compositions 
where 10\% of \Nifs\ has been replaced with \Nife\ and \Feff, while in the 
dashed ones this value is 20\%.  In all cases, light curves based on model 
Ni08 are the brightest, while those based on model Ni04 are the dimmest.}
\end{figure*}


\clearpage
\begin{figure*}
\psfig{figure=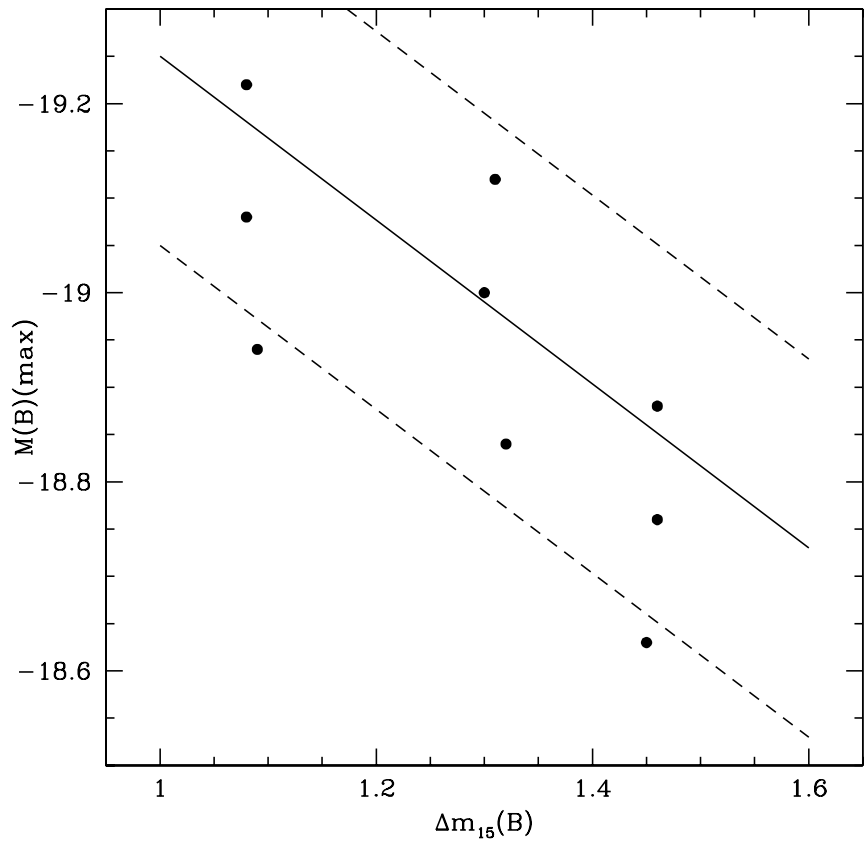,width=12cm} 
\caption{The position of the models on the brightness--decline 
rate diagram. The ridge line is the locus of the Calan-Tololo SNe 
(Hamuy et al.\ 1996a), shifted for $H_0 = 72$\,km\,s$^{-1}$\,Mpc$^{-1}$, 
and the dashed lines are the 1-$\sigma$ dispersion. }
\end{figure*}


\begin{thebibliography}{}

\newcommand{\apj}{ApJ}
\newcommand{\mnras}{MNRAS}
\newcommand{\aap}{A\&A}
\newcommand{\aj}{AJ}
\bibitem[Arnett(1982)]{arn82} Arnett, D.  1982, \apj, 253, 785

\bibitem[Arnett(1996)]{arn96} Arnett, D. 1996,  Supernovae and
    Nucleosynthesis, Princeton Univ. Press, Princeton

\bibitem[Benetti et al.(2004)]{ben04} Benetti, S., et al.\ 2004, \mnras, 348,
    261

\bibitem[Branch, Romanishin, \& Baron(1996)]{branch96} Branch, D.,
    Romanishin, W. Baron, E. 1996, \apj, 467, 473

\bibitem[Contardo, Leibundgut, \& Vacca(2000)]{cont00} Contardo, G.,
    Leibundgut, B., Vacca, W.D. 2000, \aap, 359, 876

\bibitem[Dominguez, H\"{o}flich, \& Straniero(2001)]{dom01} Dominguez, I., 
    H\"{o}flich, P., Straniero, O. 2001, \apj, 557, 279

\bibitem[Gamezo, Khokhlov, \& Oran(2005)]{gam05} Gamezo, V., Khokhlov, A.M., 
    Oran, E. 2005, \apj, 623, 337 

\bibitem[Hamuy et al.(1996a)]{ham96a} Hamuy, M., Phillips, M.M., Schommer, 
    R.A., Suntzeff, N.B., Maza, J., Aviles, R. 1996a, \aj, 112, 2391

\bibitem[Hamuy et al.(1996b)]{ham96b} Hamuy, M., Phillips, M.M.,
   Suntzeff, N.B., Schommer, R.A., Maza, J., Smith, R.C., Lira, P.,
   Aviles, R. 1996b, \aj, 112, 2438

\bibitem[Hamuy et al.(2000)]{ham00} Hamuy, M., Trager, S.C., Pinto, P.A.,
    Phillips, M.M., Schommer, R.A., Ivanov, V., Suntzeff, N.B. 2000, \aj,
    120, 1479

\bibitem[H\"oflich et al.(1996)]{hof96} H\"{o}flich, P., Khokhlov, A.M.,
    Wheeler, J.C., Phillips, M.M., Suntzeff, N.B., Hamuy, M. 1996, \apj,
    472, L81
    
\bibitem[H\"oflich et al.(1998)]{hwt98} H\"{o}flich, P., Wheeler, J.C., 
    Thielemann, F.K. 1998, \apj, 495, 617

\bibitem[Khokhlov, M\"uller, \& H\"oflich(1993)]{kmh93} Khokhlov, A.,
    M\"uller, E., H\"oflich, P. 1993, \aap 270, 223

\bibitem[Lesaffre et al.(2006)]{les06} Lesaffre, P., et al., 2006, \mnras, in
    press (astro-ph/0601443) 

\bibitem[Lucy(1999)]{lbl99} Lucy,~L.B. 1999, \aap, 345, 211

\bibitem[Mazzali(2000)]{maz00} Mazzali,~P.A. 2000, \aap, 363, 705

\bibitem[Mazzali et al.(1998)]{maz98} Mazzali, P.A., Cappellaro, E.,
   Danziger, I.J., Turatto, M, Benetti, S. 1998, \apj, 499, L49

\bibitem[Mazzali \& Lucy(1993)]{m&l93} Mazzali,~P.A., Lucy,~L.B. 1993,
    \aap, 279, 447

\bibitem[Mazzali et al.(2001)]{maz01} Mazzali, P.A.,  Nomoto, K., 
    Cappellaro, E., Nakamura, T., Umeda, H., Iwamoto, K. 2001, \apj, 547, 988

\bibitem[Nomoto et al.(1984)]{nom84} Nomoto, K., Thielemann, F.-K.,
    Yokoi, K. 1984, \apj, 286, 644

\bibitem[Nugent et al.(1995)]{nug95} Nugent, P., Phillips, M., Baron, E.,
    Branch, D., Hauschild, P.H. 1995, \apj, 455, L147

\bibitem[Pauldrach et al.(1996)]{pau96} Pauldrach, A.W.A., Duschinger, M., 
    Mazzali, P.A., et al.\ 1996, \aap, 312, 525

\bibitem[Perlmutter et al.(1997)]{perl99} Perlmutter, S., et al.\ 1997, \apj, 
    483, 565

\bibitem[Phillips(1993)]{phil93} Phillips, M.M. 1993, \apj, 413, L105

\bibitem[Phillips et al.(1999)]{phil99} Phillips, M.M., Lira, P., 
    Suntzeff, N.B., Schommer, R.A., Hamuy, M., Maza, J. 1999, \aj, 118, 1766


\bibitem[Riess et al.(1998)]{riess98} Riess, A.G., et al.\ 1998, 
    \aj, 116, 1009

\bibitem[R\"opke et al.(2005)]{roepke05} R\"opke, F., Gieseler, M., Reinecke,
    M., Travaglio, C., Hillebrandt, W. 2005, \aap, submitted (astro-ph/0506107)
        
\bibitem[Timmes, Brown, \& Truran(2003)]{tim03} Timmes, R., Brown, E.F., 
    Truran, J.W. 2003, \apj, 590, L83

\bibitem[Travaglio et al.(2005)]{travaglio05} Travaglio, C., Hillebrandt, W., 
    Reinecke, M. 2005, \aap, 443, 1007

\bibitem[van den Bergh(1997)]{vdb97} van den Bergh, S. 1997, \aj, 113, 1

\bibitem[van den Bergh, Li, \& Filippenko(2005)]{vdb05} van den Bergh, S., Li, 
    W., Filippenko, A.V. 2005, PASP, 117, 773

\bibitem[Wang, H\"oflich, \& Wheeler(1997)]{wang97} Wang, L., H\"oflich, P.,
    Wheeler, J.C. 1997, \apj, 476, L27

\bibitem[Wang et al.(2005)]{wang05} Wang, X., Wang, L., Zhou, X., Lou, Y.-Q.,
    Li, Z. 2005, \apj, 620, L87

\end{thebibliography}
\end{document}